\documentclass[aps,prc,twocolumn,groupedaddress,showpacs]{revtex4}

\usepackage{graphicx}

\begin{document}
\title{Evidence for a secondary
bow in Newton's zero-order nuclear rainbow
 }

\author{
S. Ohkubo$^{1,2}$ and  Y. Hirabayashi$^3$  
 }
\affiliation{$^1$ Research Center for Nuclear Physics, Osaka University, 
Ibaraki, Osaka 567-0047, Japan }
\affiliation{$^2$Department of Applied Science, Kochi Women's University, 5-15 Eikokuji-cho, Kochi 780-8515, Japan  }
\affiliation{$^3$Information Initiative Center,
Hokkaido University, Sapporo 060-0811, Japan}

\date{\today}

\begin{abstract}

 Rainbows are generally considered to be  caused 
 by {\it static} refraction  and reflection. A primary and a secondary rainbow  appear
 due to refraction and internal reflection in a raindrop as explained by Newton.
 The quantum nuclear rainbow, which is  generated by   refraction in the  nucleus droplet,
 only  has a ``primary'' rainbow.  Here we show for the first time  evidence for the existence of a
secondary   nuclear rainbow  generated dynamically   by coupling to an excited state
without internal reflection. 
This has been demonstrated   for experimental $^{16}$O+$^{12}$C  scattering
 using   the    coupled  channel method  with an {\it extended} double folding 
potential derived from   microscopic {\it  realistic}  wave functions  for $^{12}$C  and $^{16}$O. 
 \end{abstract}

\maketitle

Newton,  who explained the mechanism of the  rainbow by  refraction and reflection 
\cite{Newton,Nussenzveig1977,Adam2002}, believed in the existence of the zero-order rainbow without internal 
reflection in a droplet.
Newton's zero-order rainbow \cite{Bohren1991}  was  realized in quantum systems when 
Goldberg {\it et al.} \cite{Goldberg1972} observed a nuclear rainbow in  high energy 
 $\alpha$-particle scattering. The quantum nuclear rainbow is generated by  the  
  nuclear potential, which acts as a Luneburg lens \cite{Michel2002},
where refraction is the only active mechanism   and    a ``primary'' rainbow can only exist because
 there are no   higher order  terms like  in Nussenzveig's  expansion for the  meteorological
 rainbow \cite{Nussenzveig1969A}.
We report here   evidence for the unexpected existence of a  
secondary bow in the nuclear rainbow.

 The nuclear rainbow is very important in  determining the interaction 
potential family   up to  the
 internal region without discrete ambiguity
  \cite{Goldberg1972}
and has been studied extensively  \cite{Khoa2007,Anni2001}. The nuclear rainbow  is  also  
powerful for  studies of nuclear cluster structure in the  bound and unbound energy 
regions.
 In fact, the  global nuclear potential, which describes   nuclear rainbow scattering 
for  typical  $\alpha$+$^{16}$O,   $\alpha$+$^{40}$Ca  and  $^{16}$O+$^{16}$O
 systems, can reproduce scattering over a wide range of incident energies and   
 cluster structures of $^{20}$Ne,
$^{44}$Ti and $^{32}$S    in the bound and unbound energy
 regions, respectively,  in a unified way \cite{Michel1998,Ohkubo1999,Ohkubo2002}. %
  In contrast to the $^{16}$O+$^{16}$O system, the Airy structure of the nuclear rainbow  
for the asymmetric  $^{16}$O+$^{12}$C system, for which refraction is very strong,
 is  clearly observed  in  experimental angular  distributions without being obscured by
 symmetrization. For this system, a  global 
potential that  reproduces the  energy evolution of the  Airy minimum in the angular
 distributions     over a wide range of incident energies at $E_L=62\sim$1503 MeV 
 \cite{Ogloblin1998,Nicoli2000,Ogloblin2000,Szilner2001,Khoa1994}
 also successfully explains the cluster structure 
 and molecular resonances with the $^{16}$O+$^{12}$C configuration in $^{28}$Si in a unified way
 \cite{Ohkubo2004}.
 However, the global   potential was confronted with  a  serious difficulty when  a new 
 measurement of the angular  distribution in $^{16}$O+$^{12}$C scattering  at   $E_L=281$ MeV
  up to $\theta$$\approx$100$^\circ$ showed that  an  Airy  minimum of the nuclear rainbow  appears at  the 
much larger angle $\theta$$\approx$70$^\circ$ \cite{Ogloblin2003} than expected
 in the  global potential ($\theta$$\approx$45$^\circ$).
 It has been     impossible 
to model a first order  Airy  minimum $A1$ at this large angle with the established global 
 potential \cite{Ogloblin2003}.
 Thus the reason  why  the global potential 
fails  to  reproduce the Airy minimum of the nuclear rainbow at $E_L=281$ MeV
 has been a mystery 
 and   raised  a serious question of  the common belief that nuclear rainbow scattering can 
  determine the potential  uniquely.

The purpose of this paper is to  present  for the first time   evidence for the existence of a 
secondary nuclear rainbow in the  $^{16}$O+$^{12}$C  scattering  
   generated dynamically by a  quantum effect  without internal reflections  of classical concept, which
is   completely different  from the Newton's     meteorological secondary rainbow.
We show that the Airy minimum at around $\theta$$\approx$70$^\circ$,  $A1^{(S)}$, is not a
 first order 
Airy minimum, $A1^{(P)}$, of the well-known (``primary'') nuclear rainbow due to the nuclear potential 
but a first order Airy  minimum of a  {\it secondary}  bow of the nuclear rainbow. 
The difficulty of the global potential is solved by the discovery of the secondary nuclear rainbow.

We study   rainbow scattering for  $^{16}$O+$^{12}$C  with an extended double folding 
(EDF) model that describes all the diagonal and off-diagonal coupling potentials 
derived from  the microscopic  {\it realistic} wave functions for $^{12}$C  
and $^{16}$O  using  a density-dependent   nucleon-nucleon force.
  The diagonal and coupling potentials 
for the $^{16}$O+$^{12}$C system are calculated using the EDF  model
 without introducing a normalization factor:
\begin{eqnarray}
\lefteqn{V_{ij,kl}({\bf R}) =
\int \rho_{ij}^{\rm (^{16}O)} ({\bf r}_{1})\;
     \rho_{kl}^{\rm (^{12}C)} ({\bf r}_{2})} \nonumber\\
&& \times v_{\it NN} (E,\rho,{\bf r}_{1} + {\bf R} - {\bf r}_{2})\;
{\it d}{\bf r}_{1} {\it d}{\bf r}_{2} ,
\end{eqnarray}
\noindent where $\rho_{ij}^{\rm (^{16}O)} ({\bf r})$ is the diagonal ($i=j$) or transition ($i\neq j$)
 nucleon 
 density of  $^{16}$O 
  taken from  the microscopic $\alpha$+$^{12}$C  cluster model  wave functions calculated
 in  the orthogonality 
condition model (OCM) in Ref.\cite{Okabe1995}, which uses  a  realistic size parameter both 
 for the $\alpha$ particle and $^{12}$C  and is an extended version of
Ref.\cite{Suzuki1976}, which reproduces almost  all the energy levels  
well up  to $E_x$$\approx$13 MeV and the  electric transition probabilities  in $^{16}$O. 
The wave functions have been  successfully used for the systematic 
analysis of elastic and inelastic  $\alpha$+$^{16}$O scattering over a wide range of incident energies 
and the $\alpha$ cluster structure study of $^{20}$Ne \cite{Hirabayashi2013}.
$\rho_{kl}^{\rm (^{12}C)} ({\bf r})$ represents the diagonal ($k=l$) or transition ($k\neq l$)
 nucleon density of $^{12}$C which is calculated using the microscopic three $\alpha$ cluster model 
in the resonating group method \cite{Kamimura1981}. This model reproduces the 
$\alpha$ cluster, $\alpha$ condensate and shell-like structures of 
 $^{12}$C  well 
 and the  wave functions     have  been checked for many experimental
 data including charge form factors and  electric transition probabilities.  
 For the  effective interaction   $v_{\rm NN}$     we use  
 the DDM3Y-FR interaction \cite{Kobos1982}, which takes into account the
finite-range nucleon  exchange effect.
 An imaginary potential with a  
Woods-Saxon volume-type form factor is introduced   phenomenologically to take into account the effect
of absorption due to other channels.

\begin{table*}[tbh]
\begin{center}
\caption{ \label{Table I}
The   volume integral per nucleon pair $J_V$  of the 
 the ground state diagonal potential   and the   imaginary potential parameters used in the
 single channel double folding calculations in Fig.~1 and coupled channels calculations   with EDF 
in Fig.~2.
}
\begin{tabular}{ccccccccccc}
 \hline
  \hline
$E_{L}$ &  $J_V$      & $W_V$  &$R_V$ &$a_V$  & $W_V$ & $R_V$  & $a_V$   & $W_V$ & $R_V$  & $a_V$       \\    
    &         & \multicolumn{3}{c}{(DF optical model)}   &  \multicolumn{6}{c}{(Coupled channels calculation with EDF)}           \\           
    &         & \multicolumn{3}{c}{}               &  \multicolumn{3}{c}{($^{12}$C excitation only)}  \hspace{3mm}   & 
 \multicolumn{3}{c}{($^{12}$C  and $^{16}$O excitation)}   \\           
(MeV)    &(MeV fm$^3$) &(MeV) &(fm) &(fm) &(MeV)  & (fm) & (fm)   &(MeV)  & (fm) & (fm)\\   
 \hline
 \hline
 200   &  301     &  20.0   & 5.5 & 0.85   & 17.0 & 5.7 & 0.75    & 18.0  & 5.6 & 0.65 \\
 230   &  296     &  22.5 & 5.6 & 0.70   & 19.5 & 5.6 & 0.75  & 18.5   & 5.6 & 0.60   \\
 260   &  290     &  23.0  & 5.6 & 0.65   & 19.0   & 5.6 & 0.75    & 18.5   & 5.6 & 0.60 \\
 281   &  285     &  23.5   & 5.6 & 0.65 & 19.0   & 5.6 & 0.75    & 18.0   & 5.6 & 0.60   \\   
300   &  281     &  25.0   & 5.6 & 0.65 &  20.5   & 5.6 & 0.75    & 19.5   & 5.6 & 0.60   \\
 330   &  275     &  23.5   & 5.6 & 0.65   & 19 .0  & 5.6 & 0.75      & 17.0   & 5.6 & 0.60           \\
 608   &  229     &  20.0   & 5.7 & 0.60   & 17.0   & 5.8 & 0.60        & 16.0   & 5.6 & 0.60         \\
 \hline                          
 \hline                          
\end{tabular}
\end{center}
\label{Table1}
\end{table*}
\par

\begin{figure}  [t]
\includegraphics[keepaspectratio,width=7.6cm] {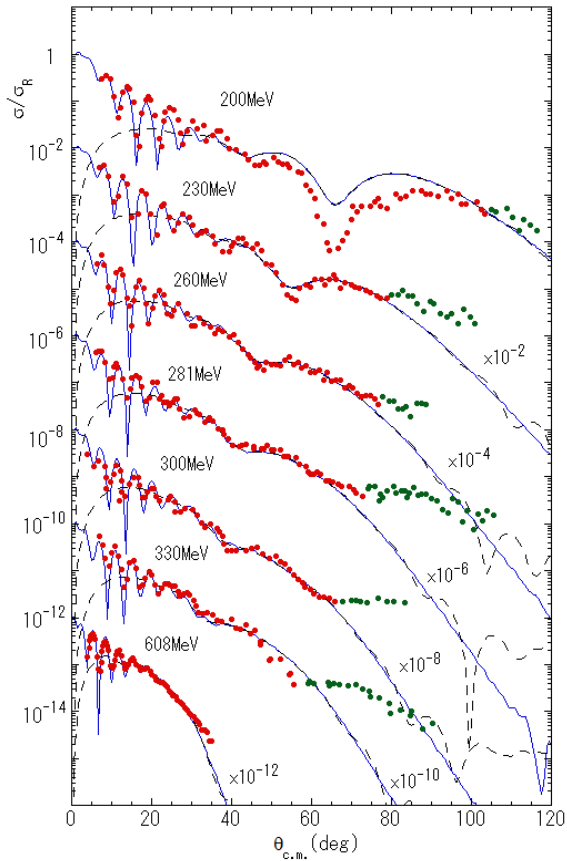}
 \protect\caption{\label{fig.1} {(Color online) 
 Comparison of the conventional single channel  DF potential calculations (blue solid line) with experimental 
angular distributions (red and green points)  in  $^{16}$O+$^{12}$C rainbow scattering
 \cite{Ogloblin2000,Brandan2001,Demyanova2004,Brandan1986}.
The long dashed line displays the calculated farside components.  
}
}
\end{figure}

\begin{figure}[b]
\includegraphics[keepaspectratio,width=7.6cm] {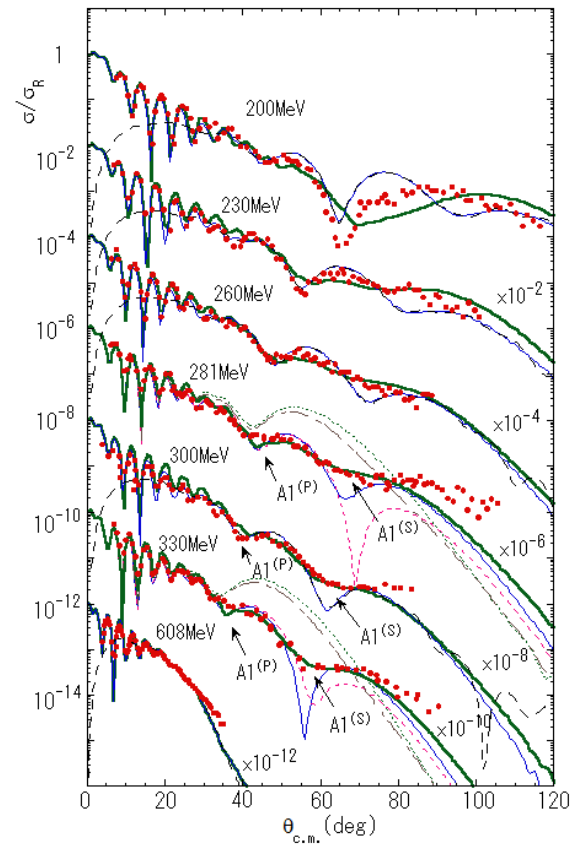}
\protect\caption{\label{fig.2} {(Color online) 
 Same as Fig.~1 but for the CC calculations.
  The calculated cross sections  with   coupling to the $2^+$ 
and $3^-$ states of $^{12}$C  (blue solid line) and  the farside components (long dashed line) are displayed in comparison with the 
experimental   data (points) \cite{Ogloblin2000,Brandan2001,Demyanova2004,Brandan1986}. 
The  calculations with  coupling to the $2^+$ state only (pink dashed line), 
$3^-$ state only (dashed dotted line) and no coupling (dotted line) are shown for
 $E_L$=281 MeV and 330 MeV.
The Airy minimum of the ``primary'' nuclear rainbow, $A1^{(P)}$,  and that of 
the secondary nuclear rainbow, $A1^{(S)}$, are indicated.
 The CC calculations including  coupling of the ground state  to the $3^-$  and 
$2^+$  states  of $^{16}$O in addition to  coupling to the $2^+$ and $3^-$ states
of $^{12}$C are displayed  by the green thick solid line (see text).
 }
}
\end{figure}

 First we show that the puzzling nuclear rainbow   for the  $^{16}$O+$^{12}$C system at
   $E_L=281$ MeV cannot be explained by the traditional view that
 a rainbow is generated  by a strong attractive nuclear potential. 
In Fig.~1, the angular distributions  of elastic $^{16}$O+$^{12}$C  scattering 
calculated  using the conventional  single channel  double folding (DF)  potential  without channel couplings  (blue solid lines)  and
 the refractive farside components (dashed lines)   are  compared with  the experimental 
  data (points)  at $E_L=200\sim$260 MeV \cite{Ogloblin2000},
 281 MeV \cite{Ogloblin2003}, 300 MeV \cite{Brandan2001}, 330 MeV \cite{Demyanova2004},
  and 608  MeV   \cite{Brandan1986}.
For the imaginary potential, the strength parameter around $20$ MeV  was found to 
fit the data  while  the radius parameter  and the diffuseness parameter 
 were fixed at  around 5.6 fm and   0.65 fm, respectively.  
 The potential parameters used and the  values of the volume integral per
 nucleon pair of the 
DF potential, $J_V$,  are given in Table I.
We found that  the  DF potential works well without introducing a normalization factor, 
  values of the volume integral per nucleon pair   are consistent 
with those used in other
 DF optical  model  calculations \cite{Nicoli2000,Ogloblin2000,Khoa1994,Brandan2001} and  the DF potential used
   belongs to the same global potential family
 found in the $E_L=62\sim$124 MeV  \cite{Nicoli2000} and  
$E_L=132\sim$1503 MeV regions \cite{Ogloblin2000,Khoa1994}.
We see that the angular distributions are dominated by the refractive farside scattering.
In Fig.~1, the discrepancies are  clear for  the data at $E=281$ MeV, 300 MeV and  
especially  330 MeV.
 It is evident  that the theoretical calculations using a conventional single channel  DF potential
 fail to reproduce the experimental data  at 281$\sim$330 MeV. 
This shows that behavior not explained by the global potential 
(visualized by green  points)  is a nuclear 
rainbow that is  not understood by the traditional concept of rainbow scattering.

  Next we show that the disagreement in Fig.~1 is not a drawback of the used global
 DF potential
 but a manifestation of the emergence of a new  rainbow which is beyond the optical potential
 model description of the nuclear rainbow in the Luneburg lens picture \cite{Michel2002}.
We calculate the  $^{16}$O+$^{12}$C  scattering  with the EDF potential 
using CC  method which takes into account  coupling to
 the excited 2$^+$ and   3$^-$ states of the  deformed   $^{12}$C at the excitation energy 
4.44 MeV and 9.64 MeV, respectively.
 The strength of the imaginary potential was slightly reduced since the channel
 coupling is explicitly introduced, 
and the  radius parameter and the diffuseness parameter were  fixed at around 5.6 fm and 0.75 fm,
 respectively (Table I).
 In Fig.~2  calculated angular distributions  are displayed in comparison with 
the experimental data.  
We see that  the  agreement between the calculations (blue solid lines)
and the experimental data is considerably
 improved  compared with Fig.~1.  (As discussed later,  the  larger oscillations than
 the data  at the large angles are reduced 
 by including additional coupling to the excited states of $^{16}$O explicitly  as displayed 
 by the green thick solid line and the fit to the  data is   further improved.) 
The calculation reproduces the new  Airy minimum of 
the nuclear rainbow in the experimental data at 281$\sim$ 330 MeV well. 
The rainbow angle, $\theta_R$, where the  fall-off of the  cross sections of  the  dark side 
of the  rainbow scattering starts   is shifted backward  in accordance with  the appearance 
of a new Airy minimum. Since the new minimum, $A1^{(S)}$,  at around $\theta=$60$\sim$70$^\circ$ 
 just before  the bump  is  caused by the farside 
 scattering (dashed lines), it is clear that the new   Airy minimum
 is produced   by the refractive scattering  with channel coupling.
Even at $E_L=200$ MeV, 230 MeV and 260 MeV the   agreement of the calculated  angular 
distribution with the experimental data is  considerably improved  in the regions
 $\theta > 100$$^\circ$,   $\theta=80^\circ\sim100^\circ$, and 
$\theta=70^\circ\sim90^\circ$, respectively.

As shown at $E_L=$281 MeV and 330 MeV in Fig. 2,  in creating a new Airy minimum, $A1^{(S)}$,  at the larger angles 
  the coupling to the $2^+$ state  is  dominant, while the effect of the 
$3^-$ state   is negligible and  can be simply  incorporated into the imaginary  potential. 
Refraction via  coupling to the   $2^+$ state plays the role  of a dynamical secondary lens 
which is completely different from  the ordinary static lens of the nuclear potential interpreted 
as a  Luneburg lens \cite{Michel2002}. 
 The effect of the dynamical lens is not possible  to incorporate   into the  DF optical model potential 
as a normalization  constant which is often used in the conventional DF model approach  \cite{Brandan1997}. 
For this reason   the global conventional DF optical model  potential failed to 
reproduce the Airy structure in the 281$\sim$330 MeV data in Fig.~1.

\begin{figure}[tbh]
\includegraphics[keepaspectratio,width=7.6cm] {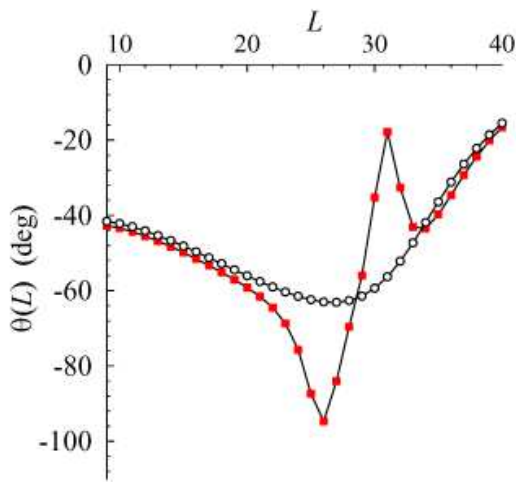}
 \protect\caption{\label{fig.3} {(Color online) 
The deflection  functions calculated from the real DF optical model   (circle)  and 
the CC method (square) with the EDF potential at $E_L=330$ MeV are compared. 
The line is to guide the eye.
}
}
\end{figure}

By investigating the deflection function we  show that this new Airy minimum,
 $A1^{(S)}$,  of
 the nuclear rainbow in the range 281$\sim$330 MeV  is due to  a {\it secondary nuclear rainbow},
which is dynamically  generated  in the classically forbidden angular region, i.e., in the 
 dark side of the ordinary  ``primary'' nuclear rainbow.
In  Fig.~3 the deflection functions  determined  from the phase shifts calculated
 in the CC  method  and in the conventional DF optical model without imaginary potential  
 are displayed  at 330 MeV. 
In the conventional DF optical model calculation we see  a single minimum
 at the orbital angular momentum $L_R=27$  with the  rainbow angle  $\theta_{R}=63^\circ$
  consistent with 
  the textbook fact that the nuclear rainbow  is caused   by a refraction in the    nuclear attractive potential \cite{Newton1966}.
On the other  hand, in the CC calculation  the deflection
 function is drastically changed to have  more than two extrema, which
   means the existence of more than two nuclear rainbows in elastic scattering
by  a new mechanism. The investigation of the $S$-matrix shows that the 
coupling to the $2^+$ state is very strong for the relevant $L=20\sim35$ region, especially at the minima.
The two extrema at $L_{R0}=31$ with $\theta_{R0}=18^\circ$  and at $L_{R1}=34$
 with the rainbow angle $\theta_{R1}=43^\circ$    are  located at forward angles, 
 while the  minimum   with  the largest rainbow angle $\theta_{R2}=95^\circ$
appears at  $L_{R2}=26$. 
In contrast to the meteorological rainbow, the bright sides of the two rainbows where 
  the Airy maximum and minimum  appear 
 with $\theta_{R1}$ (``primary'' nuclear rainbow) and $\theta_{R2}$
 (secondary nuclear rainbow) stand on the same side of the nuclear rainbow. Therefore
the dark side of  the ``primary'' nuclear rainbow where  the cross sections fall toward 
large angles in the angular distribution  is  masked by 
the bright side of the secondary nuclear rainbow   and difficult to observe.  
 The experimental Airy minimum at the large angles, $A1^{(S)}$,  and the  fall-off  are due to this  
secondary nuclear rainbow
and it is  reasonable that the conventional global  DF optical model  with a single deflection minimum,
 which creates an Airy minimum, $A1^{(P)}$, of the ``primary'' nuclear rainbow due to the
 nuclear potential,  cannot reproduce the  Airy
 minimum of the secondary nuclear rainbow at 281$\sim$300 MeV. The precursor of the secondary nuclear 
rainbow  can already be  seen in the  poor agreement of the calculated angular distributions
 in the conventional DF optical  model  at 260 MeV in Fig.~1.  
   It is  startling that a secondary 
nuclear rainbow appears  for the nuclear system where the active mechanism is only refractive.
This has not been anticipated  from the  meteorological  secondary rainbow 
for which  reflection is involved
 \cite{Newton,Adam2002,Nussenzveig1969A,Nussenzveig1977,Newton1966}.
 A dynamical second lens is only possible in a quantum system
 such as the $^{16}$O+$^{12}$C scattering as demonstrated in this paper.
The traditional view that  the rainbow is generated  by {\it static} refraction 
 and reflection of  classical concept  is changed  in nuclear  context.

 In Fig.~4  calculated results without  and with the imaginary potentials
 are compared.  We can confirm that  the Airy minimum $A1^{(S)}$,  
 which appears also in the absence of  the imaginary potential,
  is entirely due to  the real  potential and is created by coupling to the $2^+$ 
state of $^{12}$C.

\begin{figure}[tbh]
\includegraphics[keepaspectratio,width=7.6cm] {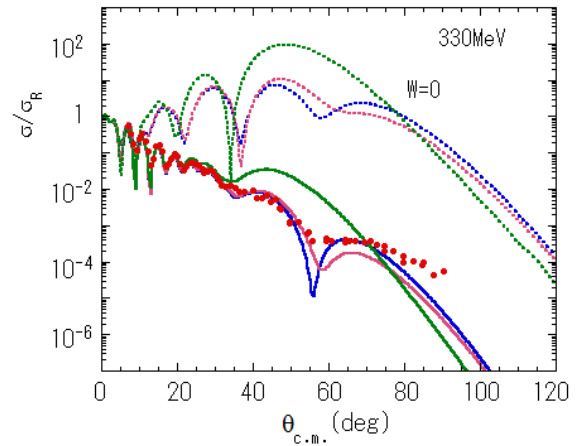}
 \protect\caption{\label{fig.4} {(Color online) 
 Angular distributions in  $^{16}$O+$^{12}$C  scattering at $E_L$=330 MeV
 calculated in the single channel (green solid line) and 
CC method (blue and pink solid line) with  the imaginary potential
 in Table I are compared with those calculated by switching off the imaginary potential
 (dotted line) and the experimental data (points) \cite{Demyanova2004}.
 The blue  and    pink  lines
represent the CC results with    coupling to the $2^+$ 
and $3^-$ states of $^{12}$C and those with
 coupling to the $2^+$ state of $^{12}$C only, respectively.
}
}
\end{figure}

We investigate  the  effects of the channels  coupling to the excited states 
  of  $^{16}$O on the creation of the Airy minimum $A1^{(S)}$.
In Fig.~2    the angular distributions calculated  including the channels coupling of the ground 
state  to the  excited states, 3$^-$ (6.13 MeV)  and 2$^+$ (6.92 MeV)   of  $^{16}$O in 
addition   to the excited states of  $^{12}$C are displayed by the green thick solid line. 
Potential parameters used are given in Table I.  We see that the large oscillations at large angles 
seen in the  blue lines  are suppressed and 
   the agreement  with the experimental  data   is improved further especially at the large angles  
around and beyond the Airy minimum.  This coupling  tends  to obscure the deep Airy minimum
  created by  coupling to the 2$^+$ state  of $^{12}$C  as an imaginary potential generally  does.
These results show that the Airy minimum $A1^{(S)}$  is  due to the  2$^+$ state  of $^{12}$C.
Our CC calculations using the   imaginary potentials  interpolated from
  $E_L=$330 MeV   and 608 MeV  predict  the  persistency of the Airy minimum of the nuclear
 rainbow   at the higher energies  between $E_L=$350 MeV and 500 MeV. 
Especially  at around $E_L=$400 MeV a clear deep Airy minimum, which is hardly obscured by the coupling
 to the excited states of $^{16}$O,  is predicted.  The experimental
 observation of the Airy minimum at these energies is expected.

Finally we will make a brief discussion on the contribution of the  transfer reaction channels.
 As for the elastic transfer of $\alpha$ particles \cite{Szilner2002},
we see  in the recent    coupled reaction channel   calculations of 
Ref.\cite{Rudchik2010}
 that the exchange of the $\alpha$ particle (elastic transfer)  is   three order
 of magnitude smaller than the elastic scattering cross sections.
 Although a complete description of the data may require both
inelastic and exchange couplings,
as  for the  one nucleon exchange effect, which is suggested to 
  prevail over other transfer reactions to
affect the Airy minimum $A1^{(S)}$ for the 230 MeV data
 in Ref.\cite{Rudchik2010},
 our calculations  take into account it by using the effective  interaction 
DDM3Y, in which   the knock-on exchange   effect is incorporated 
\cite{Kobos1982,Brandan1997}.

To summarize, we have shown for the first time the evidence for the existence of a secondary
 bow in the Newton's zero-order nuclear rainbow  by analyzing $^{16}$O+$^{12}$C scattering 
 using a coupled channel   method with an extended  
 double folding (EDF) potential derived from the microscopic {\it  realistic}  wave functions 
for $^{12}$C and $^{16}$O.
The secondary nuclear rainbow  appears dynamically in the classically forbidden angular 
region  of the dark side of the ordinary ``primary'' nuclear rainbow  due to the creation of a
new rainbow  angle   caused by  strong coupling to the $2^+$ state of $^{12}$C.
 This  plays the role of  a
second lens in addition to the static  lens caused by the nuclear potential.
The traditional view that  the rainbow is understood by the  classical concept  of   refraction 
 and reflection is   changed  in nuclear  context. 
 The discovery of a secondary rainbow solves the dilemma concerning the failure  of the 
 global potential  for the  $^{16}$O+$^{12}$C system.
 It is  desired to observe a secondary rainbow  in other systems.

One of the authors (SO) thanks the Yukawa Institute for Theoretical Physics for
 the hospitality extended  during a stay in February 2013. 
Part of this work was  supported by the Grant-in-Aid for the Global COE Program ``The Next
 Generation of Physics, Spun from Universality and Emergence'' from the Ministry 
of Education, Culture, Sports, Science and Technology (MEXT) of Japan.


\begin{thebibliography}{aa}

\bibitem {Newton}
I.  Newton, {\it Opticks or, a Treatise of the Reflexions, Refractions, Inflexions and 
Colours of Light.}  (London, 1704); (Dover publications, New York, 1952).

\bibitem {Nussenzveig1977}
H. M.  Nussenzveig,
Sci. Am. {\bf 236}, 116 (1977).
\bibitem {Adam2002}
 J. A. Adam,
Phys. Rep. {\bf 356}, 229 (2002).
\bibitem {Bohren1991}
C. F. Bohren and  A. B.  Fraser, 
Am. J. Phys. {\bf 59}, 325 (1991).
\bibitem {Goldberg1972}
D. A. Goldberg, S. M.  Smith, 
 Phys. Rev. Lett. {\bf 29}, 500 (1972);
 D. A. Goldberg, S. M. Smith, G. F.   Burdzik, 
 Phys. Rev. C {\bf 10}, 1362 (1974).
\bibitem {Michel2002}
F. Michel, G. Reidemeister, S.   Ohkubo, 
Phys. Rev. Lett. {\bf 89}, 152701 (2002).
\bibitem {Nussenzveig1969A}
H. M. Nussenzveig,
 J. Math. Phys. {\bf 10}, 82 (1969);
J. Math. Phys. {\bf 10}, 125 (1969).
\bibitem {Khoa2007} 
D. T. Khoa,  W. von Oertzen, H.G. Bohlen, S.  Ohkubo, 
J. Phys. {\bf G 34}, R111 (2007).
\bibitem {Anni2001} 
 R. Anni,  Phys. Rev. C {\bf 63}, 031601(R) (2001).
\bibitem {Michel1998}
F. Michel,  S.  Ohkubo, G.  Reidemeister, 
   Prog. Theor. Phys. Suppl. {\bf 132}, 7 (1998). 
\bibitem {Ohkubo1999}  
S. Ohkubo,  T.  Yamaya, and  P. E.   Hodgson,
 Nuclear clusters. in {\it Nucleon-Hadron Many-Body
 Systems}, 
(eds. H. Ejiri  and H. Toki, ) p. 150  (Oxford University Press, Oxford, 1999).
\bibitem {Ohkubo2002} 
 S. Ohkubo and  K.  Yamashita, 
 Phys. Rev. C {\bf 66}, 021301(R) (2002).
\bibitem {Ogloblin1998}
 A. A. Ogloblin {\it et al.},
 Phys. Rev. C {\bf 57}, 1797 (1998).
\bibitem {Nicoli2000}
M. P. Nicoli  {\it et al.}, 
Phys. Rev. C  {\bf  61},  034609 (2000).
\bibitem {Ogloblin2000}
A. A. Ogloblin {\it et al.}, 
 Phys. Rev. C {\bf 62}, 044601 (2000).
\bibitem {Szilner2001}
S. Szilner  {\it et al.}, 
 Phys. Rev. C {\bf  64}, 064614  (2001).
\bibitem{Khoa1994} 
D. T. Khoa, W. von Oertzen,  and  H. G.   Bohlen, 
 Phys. Rev. C  {\bf  49}, 1652  (1994).

\bibitem {Ohkubo2004} 
S. Ohkubo and  K.  Yamashita,  
 Phys.  Lett. {\bf B578}, 304 (2004).
\bibitem {Ogloblin2003}
 A. A. Ogloblin {\it et al.}, 
 Phys. At. Nucl. {\bf 66}, 1478 (2003).

\bibitem {Okabe1995} 
S.  Okabe, 
{\it Tours Symposium on Nuclear Physics II}, edited
by H. Utsunomiya {\it et al.} (World Scientific, Singapore, 1995),
p. 112.
\bibitem {Suzuki1976} 
Y.~Suzuki,
Prog. Theor. Phys. {\bf 55}, 1751 (1976);
 Prog. Theor. Phys. {\bf 56}, 111 (1976). 
\bibitem {Hirabayashi2013}
   Y.~Hirabayashi and S.~Ohkubo,
  Phys. Rev. C {\bf 88}, 014314 (2013).
\bibitem{Kamimura1981}
 M. Kamimura,
Nucl. Phys. {\bf A351}, 456 (1981).

\bibitem {Kobos1982}
A. M. Kobos {\it et al.},
 Nucl. Phys. {\bf A384}, 65 (1982);
A. M. Kobos {\it et al.},
Nucl. Phys. {\bf A425}, 205 (1984).
\bibitem{Brandan2001} 
M. E. Brandan  {\it et al.}, 
Nucl. Phys. {\bf A688}, 659 (2001).
\bibitem {Demyanova2004} 
A. S. Dem'yanova {\it et al.},
IAEA Database exfor. 
 \bibitem {Brandan1986}
 M. E.  Brandan {\it et al.},
 Phys. Rev. C {\bf 34}, 1484  (1986).

\bibitem {Brandan1997}
 M. E.  Brandan and   G. R.  Satchler,
Phys. Rep. {\bf 285}, 143  (1997). 

\bibitem {Newton1966}
R. G.   Newton, {\it Scattering Theory of Waves and Particles}
(McGraw-Hill Book Company, New York, 1966).
\bibitem {Szilner2002}
S. Szilner {\it et al.}, 
Eur. Phys. J. {\bf A 13}, 273 (2002).
\bibitem {Rudchik2010}
A. T. Rudchik {\it et al.}, 
Eur. Phys. J. {\bf A  44}, 221 (2010).

\end{thebibliography}
\end{document}